\documentclass[conference]{IEEEtran}
\IEEEoverridecommandlockouts

\usepackage{cite}
\usepackage{amsmath,amssymb,amsfonts}
\usepackage{algorithmic}
\usepackage{graphicx}
\usepackage{textcomp}
\usepackage{xcolor}
\usepackage{booktabs}
\usepackage{multirow}
\usepackage{balance}
\def\BibTeX{{\rm B\kern-.05em{\sc i\kern-.025em b}\kern-.08em
    T\kern-.1667em\lower.7ex\hbox{E}\kern-.125emX}}
\begin{document}
\title{Channel-Adaptive Region Adjacency Graph Carriers\\
for Semantic Image Communication
\thanks{This work is supported by the American University of Beirut University Research Board (URB) and Vertically Integrated Projects (VIP) program.
\par\scriptsize
\textcopyright{} 2026 IEEE. Personal use of this material is permitted. Permission from IEEE must be obtained for all other uses, in any current or future media, including reprinting/republishing this material for advertising or promotional purposes, creating new collective works, for resale or redistribution to servers or lists, or reuse of any copyrighted component of this work in other works.}
}

\author{
\IEEEauthorblockN{Karim Abdallah, Maria Slim, Mariette Awad, Hadi Sarieddeen}
\IEEEauthorblockA{
Department of Electrical and Computer Engineering \\
American University of Beirut, Beirut, Lebanon \\
Email: \{kka12, mas194\}@mail.aub.edu, 
\{mariette.awad, hadi.sarieddeen\}@aub.edu.lb
}
}
\maketitle
\thispagestyle{empty}

\begin{abstract}

Semantic image communication seeks to preserve task-relevant scene structure under limited channel resources, but carriers are often dense latent tensors or grid-aligned semantic layouts that do not explicitly encode region-level relations. This work introduces a segmentation-derived region adjacency graph (RAG) carrier, termed channel-adaptive RAG (CA-RAG), for joint source--channel coding-style image communication. Nodes store interpretable region attributes, edges preserve adjacency, channel-adaptive graph simplification (CGS) controls the node budget, and semantic belief propagation refines noisy graph embeddings before diffusion-based reconstruction. On Cityscapes, pre-channel RAG payloads are several times smaller than compressed class-index layouts in a 2,000-image study. Under additive white Gaussian noise at signal-to-noise ratios from 0 to 15~dB, CA-RAG reports higher semantic consistency than deep joint source--channel coding and a same-decoder layout baseline, with comparable perceptual quality. At 10~dB, the full-budget rate-sweep point reaches mean intersection over union (mIoU) $\!=\!0.329$ at $\!\approx\! 3.3\!\times\! 10^3$ channel uses, while the default adaptive-CGS setting reports mIoU $\!=\!0.294$ at $\!\approx\! 2.6\!\times\! 10^3$ channel uses.

\end{abstract}


\section{Introduction}

Semantic communication aims to preserve task-relevant meaning under channel and bandwidth constraints rather than uniformly protecting every source bit \cite{qin2021semanticcommunicationsprincipleschallenges, ismail2026semantic}. For image transmission, this objective is relevant in noisy and bandwidth-limited settings, where dense pixel-level transfer is inefficient. Recent systems have shown that communication pipelines can be optimized around semantic objectives, from deep learning enabled semantic communication (DeepSC) \cite{9398576}, robust semantic communication \cite{alhaj2026signdeepsc}, and deep joint source--channel coding (DeepJSCC) \cite{8723589} to diffusion-guided reconstruction with generative semantic communication (GESCO) \cite{Grassucci_2026}.

However, existing semantic image representations are often dense learned tensors or grid-aligned semantic layouts, which do not explicitly represent the transmitted semantic entities and their spatial relations. This motivates a region-level graph representation in which image regions form semantic entities and adjacency relations capture scene structure. The key insight is that such a representation enables control of the semantic transmission rate through node selection, allowing the number of transmitted regions to adapt to channel conditions.

Building on region adjacency graphs (RAGs) used in image analysis~\cite{841950}, we use segmentation-derived graphs as semantic communication carriers. Their embeddings are transmitted as analog real-valued symbols in a joint source--channel coding (JSCC)-style pipeline without a separate digital channel code. This choice has three consequences: (i) scene complexity is encoded explicitly as a count of transmitted nodes rather than implicitly inside a fixed-size tensor; (ii) region-level adjacency relations are preserved in the carrier, enabling structured receiver-side refinement; and (iii) representation size depends on region count; the primary channel-use metric excludes topology and other metadata. These properties position the proposed pipeline between feature-level systems such as DeepJSCC, which transmit dense latents, and layout-level systems such as GESCO, which transmit grid-aligned maps.

Channel-adaptive graph simplification (CGS) uses a heuristic node budget driven by the signal-to-noise ratio (SNR). Semantic belief propagation (SBP) refines graph embeddings.
The main contributions of this work are:
\begin{enumerate}
    \setlength{\itemsep}{0pt}
    \setlength{\parsep}{0pt}
    \setlength{\topsep}{2pt}
    \item We introduce RAGs as semantic communication carriers whose representation size is governed by scene-region count rather than pixel-grid size.
    \item We develop a CGS rule that controls the transmitted node budget as a function of SNR, enabling semantic rate adaptation.
\end{enumerate}

We evaluate the carrier under additive white Gaussian noise (AWGN), idealized Rayleigh equalization, and structured node-erasure stress tests, with explicit channel-use analysis and a 10~dB rate comparison against a same-decoder layout baseline. Ablations on graph encoding, SBP receiver-side refinement, CGS sensitivity, and learned-budget selection clarify component effects and motivate future work.

\section{Proposed CA-RAG Framework}

Throughout, $H\times W$ is the image size, $K$ is the class count, $N$ is the initial node count, $M$ is the retained-node count, $d$ is the embedding width, $\gamma_{\mathrm{dB}}$ is the channel SNR, $P$ is signal power, and $\sigma^2$ is noise variance. 
Bold lowercase, bold uppercase, and calligraphic symbols denote vectors, matrices or image arrays, and sets or graphs, respectively. The five pipeline stages are RAG construction, graph encoding, CGS, noisy transmission, and receiver-side graph refinement and graph-conditioned reconstruction.

\subsection{Relation to GESCO}

GESCO~\cite{Grassucci_2026} transmits a grid-aligned semantic layout and uses diffusion-based reconstruction at the receiver. We adopt diffusion-based semantic reconstruction as a receiver-side design choice and use a GESCO-inspired layout implementation as our principal baseline. 
In contrast, CA-RAG transmits region-level nodes and adjacency relations, adapts the transmitted node budget through CGS, and applies graph-structured receiver refinement through SBP. Accordingly, our claimed contribution is the graph carrier and its rate-control and refinement mechanisms, rather than the diffusion reconstruction architecture.

\subsection{RAG-Based Semantic Representation}

A segmentation model $\mathcal{S}$ maps the input image $\mathbf{X} \!\in\! \mathbb{R}^{H \times W \times 3}$ to the pixel-wise map $\mathbf{M}\!=\!\mathcal{S}(\mathbf{X})\in\{1,\dots,K\}^{H\times W}$. Its connected components define regions $\{\mathcal{R}_i\}_{i=1}^{N}$; Cityscapes instance IDs, when available, separate same-class object instances.
The RAG $\mathcal{G}\!=\!(\mathcal{V},\mathcal{E})$ has nodes $\mathcal{V}\!=\!\{v_i\}_{i=1}^{N}$, with $v_i$ representing $\mathcal{R}_i$, and edges $(i,j)\!\in\!\mathcal{E}$ between regions sharing horizontal or vertical pixel boundaries.

Each node carries a $(K\!+\!28)$-dimensional interpretable feature vector (47 dimensions for Cityscapes, where $K\!=\!19$), 
\begin{equation}
    \mathbf{f}_i = \left[\mathbf{e}_{c_i}^\top,\; a_i,\; x_i^{(c)},\;  y_i^{(c)},\; \mathbf{u}_i^\top,\; d_i\right]^\top,
\end{equation}
where $c_i$ is the region class, $\mathbf{e}_{c_i}$ its one-hot code, $a_i\!=\!|\mathcal{R}_i|/(HW)$, and $(x_i^{(c)},y_i^{(c)})$ is the centroid (row/$H$, column/$W$). $\mathbf{u}_i$ has eight bins per red--green--blue (RGB) channel, divided by region pixel count; $d_i$ is the mean Cityscapes disparity divided by 2047 (zero if unavailable).
This low-dimensional interpretable representation keeps the transmitted payload tied to a small, auditable set of scene statistics.

Excluding metadata, the layout has $HWK$ scalars; RAG features have $N(K+28)$, encoded to $Nd$ and pruned to $Md$. Since $N \ll H \cdot W$ in our experiments (Table~\ref{tab:tx_cost}), the graph carrier shifts representation size from pixel-grid resolution to scene complexity. This is a representational scaling statement rather than a formal rate-distortion claim; channel-rate evidence appears in Section~\ref{sec:ratematched}.

\subsection{Graph Encoder and Channel-Adaptive Simplification}

The graph encoder is a three-stage graph attention network (GAT) whose first two stages use four attention heads and whose final stage uses one, with hidden width \(d=64\)~\cite{velickovic2018graphattentionnetworks}. It maps $\mathbf{f}_i$ to latent node embeddings $\mathbf{z}_i\in\mathbb{R}^{64}$ while aggregating neighborhood information. CGS adapts the transmitted graph complexity to the channel state $\gamma_{\mathrm{dB}}$ via a keep-ratio rule. This linear-in-dB rule is heuristic.
CGS acts as a rate-control mechanism rather than a semantic optimizer. Let
\begin{equation}
    \alpha(\gamma_{\mathrm{dB}}) = \mathrm{clip}\!\left(\gamma_{\mathrm{dB}}/\gamma_{\mathrm{ref}},\,0,\,1\right), \quad \gamma_{\mathrm{ref}} = 20~\mathrm{dB},
\end{equation}
where $\mathrm{clip}(\cdot,0,1)$ bounds its argument to $[0,1]$ and $\gamma_{\mathrm{ref}}$ is the saturation reference. The target keep-ratio is $r(\gamma_{\mathrm{dB}}) = r_{\min} + \alpha(\gamma_{\mathrm{dB}})(r_{\max} - r_{\min})$, with defaults $r_{\min}=0.5$ and $r_{\max}=1.0$. The node budget is $M(\gamma_{\mathrm{dB}}, N) = \max(3, \operatorname{round}(r(\gamma_{\mathrm{dB}})N))$. A per-node importance score combines a multilayer perceptron (MLP) output, region size, and inverse-frequency class weighting:
\begin{equation}
    s_i = \mathrm{MLP}(\mathbf{z}_i,\,\mathbf{e}_{\mathrm{task}}) + \lambda_a \log(1+a_i) + \lambda_c\, w(c_i),
\end{equation}
where $\mathbf{e}_{\mathrm{task}}$ is a shared learned conditioning vector formed by averaging three trainable 16-dimensional embeddings, $\lambda_a$ and $\lambda_c$ weight area and class, and $w(c_i)$ is inverse class frequency. The top-$M$ nodes and their edges form $\tilde{\mathcal{G}}=(\tilde{\mathcal{V}},\tilde{\mathcal{E}})$; a discarded region with surviving neighbors merges into the one with highest importance score $s_j$, using area-weighted embedding averages.
Top-$M$ selection is optimal only under an additive importance surrogate; the full pipeline is non-linear, so we treat CGS as a heuristic rate controller and study its sensitivity in Section~\ref{sec:cgs_sensitivity} rather than claiming optimality.

\subsection{Channel Model, SBP, and Reconstruction}

The retained embeddings $\tilde{\mathbf{Z}}\in\mathbb{R}^{M\times d}$ are power-normalized and transmitted over a JSCC-style AWGN channel:
\begin{equation}
\begin{aligned}
    \mathbf{Z}_{\mathrm{tx}} &= \sqrt{\frac{MdP}{\max(\|\tilde{\mathbf{Z}}\|_F^2,10^{-8})}}\,\tilde{\mathbf{Z}},\\
    \mathbf{Y} &= \mathbf{Z}_{\mathrm{tx}} + \mathbf{Q}, \qquad q_{i\ell} \overset{\mathrm{i.i.d.}}{\sim} \mathcal{N}(0,\sigma^2),
\end{aligned}
\end{equation}
where $\|\cdot\|_F$ denotes the Frobenius norm; the floor prevents division by zero. The transmit, received, and noise matrices $\mathbf{Z}_{\mathrm{tx}}$, $\mathbf{Y}$, and $\mathbf{Q}=[q_{i\ell}]$ are $M\times d$.
Each scalar is one real channel use; $P=1$ and $\sigma^2 = P\cdot 10^{-\gamma_{\mathrm{dB}}/10}$. No entropy coding, quantization, or digital channel code is applied in the main experiments; this follows analog DeepJSCC \cite{8723589} and DeepSC \cite{9398576}. Channel uses count $Md$ embedding scalars only; Section~\ref{sec:cost} details the metadata exclusions.

For fading, $\mathbf{A}=\operatorname{diag}(g_1,\ldots,g_M)\in\mathbb{R}^{M\times M}$ has gains on its diagonal and zeros elsewhere; $\mathbf{A}\mathbf{Z}_{\mathrm{tx}}$ scales row $i$ by $g_i$ through ordinary matrix multiplication. Real gains are sampled with Rayleigh scale $1/\sqrt{2}$, matching the magnitude distribution of a zero-mean, unit-power circularly symmetric complex Gaussian coefficient; thus $\mathbb{E}[g_i^2]=1$; $\mathbb{E}[\cdot]$ denotes expectation. We test independent per-node gains (node-block) and common $g_i=g$ (graph-block); Table~\ref{tab:rayleigh} reports both.
Assuming perfect receiver channel state information (CSI),
\begin{equation}
\begin{aligned}
    \mathbf{Y}_{\mathrm{f}} &= \mathbf{A}\mathbf{Z}_{\mathrm{tx}}+\mathbf{Q}_{\mathrm{f}},\\
    \mathbf{Y}_{\mathrm{eq}} &= \bar{\mathbf{A}}^{-1}\mathbf{Y}_{\mathrm{f}}, \qquad
    \bar{\mathbf{A}}=\operatorname{diag}(\bar g_1,\ldots,\bar g_M).
\end{aligned}
\end{equation}
Here, $\mathbf{Q}_{\mathrm{f}}=[q_{\mathrm{f},i\ell}]$, with $q_{\mathrm{f},i\ell}\overset{\mathrm{i.i.d.}}{\sim}\mathcal{N}(0,\sigma^2)$, and $\bar g_i=\max(g_i,10^{-3})$. The floor limits inverse gains to $10^3$ for numerical stability, modifying ideal zero-forcing below $10^{-3}$. 
Entrywise, $y_{\mathrm{f},i\ell}=g_i z_{\mathrm{tx},i\ell}+q_{\mathrm{f},i\ell}$ and $y_{\mathrm{eq},i\ell}=y_{\mathrm{f},i\ell}/\bar g_i$. Setting all $g_i=1$ gives $\mathbf{A}=\mathbf{I}_M$ and recovers AWGN for all $d$ columns. Deep fades amplify noise.

At the receiver, SBP refines the noisy embeddings through graph-structured message passing,
\begin{equation}
    \mathbf{h}_i^{(t+1)} = (1-\delta_t)\, \mathbf{h}_i^{(0)} + \delta_t\, f_{\mathrm{upd}}\!\left(\big[\mathbf{h}_i^{(t)},\,{\textstyle\sum}_j \mathbf{m}_{ij}^{(t)}\big]\right).
\end{equation}
We initialize $\mathbf{h}_i^{(0)}$ from its received embedding. At iteration $t$, the message from neighbor $j$ to $i$ is
\begin{equation}
    \mathbf{m}_{ij}^{(t)}=f_{\mathrm{msg}}^{(t)}\!\left([\mathbf{h}_i^{(t)},\mathbf{h}_j^{(t)},\mathbf{b}_{ij}]\right),
\end{equation}
where brackets denote concatenation. The shared edge vector $\mathbf{b}_{ij}=\mathbf{b}_{ji}\in\mathbb{R}^{3}$ stores boundary length normalized by the graph's maximum and two normalized-centroid offsets. It is computed once per undirected edge and reused in both directions without sign reversal. Each $f_{\mathrm{msg}}^{(t)}$ is a learned $(2d+3)\to d\to d$ MLP with a hidden ReLU. The sum spans $i$'s neighbors; $f_{\mathrm{upd}}$ is a two-layer MLP with layer normalization. 
The damping is $\delta_t=\operatorname{sigmoid}(\tilde{\delta}_t)$, with unconstrained learned parameter $\tilde{\delta}_t$. SBP is a small message-passing refinement module inspired by belief propagation \cite{779343.779352}; its effect is empirically modest in AWGN (Section~\ref{sec:ablations}). The refined embeddings are rendered into a coarse spatial control signal and fed to a graph-conditioned diffusion reconstructor through a ControlNet branch \cite{10377881,9878449}.

\section{Experimental Setup}

The main evaluation uses the 500-image Cityscapes validation set \cite{Cordts_2016_CVPR} at $512\times 1024$, with four independent channel-noise realizations per image, yielding 2,000 image-realization pairs. Semantic maps are predicted by a pretrained SegFormer-B2, a transformer-based semantic segmentation model \cite{NEURIPS2021_64f1f27b}. The graph encoder is a three-layer GAT with $d=64$, and the reconstructor is a ControlNet-guided latent diffusion model based on Stable Diffusion 1.4 (SD-1.4) \cite{10377881,9878449}. SNRs are drawn from $\{0,5,10,15\}$~dB unless noted. All transmissions are analog (real-valued symbols, no digital channel code or quantizer). The CGS weights $\lambda_a$ and $\lambda_c$ are trainable parameters initialized to $1.0$.
The bandwidth ratio is defined as the number of channel uses per source pixel; for a retained graph with $M$ nodes and embedding width $d$, $\rho = Md/(HW)$. 

CA-RAG is trained on all 2,975 Cityscapes training images using AdamW, with learning rates $10^{-4}$ for the graph modules and renderer and $5\times10^{-5}$ for the trainable ControlNet blocks. Training SNR is sampled uniformly from $[0,20]$~dB. SegFormer-B2, the SD-1.4 variational autoencoder (VAE), and the U-Net remain frozen; the trainable modules use diffusion noise-prediction mean squared error (MSE).
At inference, the receiver uses five SBP iterations, classifier-free guidance scale $5.0$, and 20 denoising diffusion implicit model (DDIM) steps at 0--15~dB (25 at 20~dB).

We consider two baselines. DeepJSCC \cite{8723589} is a paper-faithful five-layer parametric rectified linear unit (PReLU) convolutional neural network (CNN) autoencoder, trained on Cityscapes at $256\times 512$ with SNR-aware training over $\{1,5,10,15,20\}$~dB, $C=16$ bottleneck channels, and bandwidth ratio $\rho\approx 0.021$.
It is trained for 30 epochs with Adam and MSE (batch size eight; initial learning rate $10^{-4}$). Because DeepJSCC is trained at $256\times 512$ and our method runs at $512\times 1024$, DeepJSCC is used as a reference rather than a strictly matched baseline. The GESCO-inspired layout baseline follows the transmission logic of \cite{Grassucci_2026}: the segmentation map is one-hot encoded at $256\times 512$ for present classes, power-normalized, transmitted over AWGN, denoised with our GESCO-inspired fast-denoising semantic (FDS) implementation using $5\times5$ average pooling, $5\times5$ max pooling, and a $0.5$ threshold, colorized, and fed to \emph{the same ControlNet+diffusion decoder} used by our method. This partially controls for the decoder, but the conditioning signals differ (graph-rendered control vs.\ grid layout), so the carrier is not fully isolated. We emphasize the RAG-versus-layout carrier comparison.

Reconstruction quality is measured with learned perceptual image patch similarity (LPIPS) and Fr\'echet inception distance (FID), with lower values indicating better quality for both. Mean intersection over union (mIoU) compares SegFormer-B2 predictions on original and reconstructed images, averaging over classes present in either map.
The main results use all 500 validation images with four channel-noise realizations per image. This repeated-realization protocol reduces sensitivity to a single channel draw, but FID remains sensitive to sample composition and generative artifacts, so we treat it as auxiliary evidence and base quality claims primarily on LPIPS and mIoU. Claims are restricted to this Cityscapes protocol.

\section{Results}

Tables~\ref{tab:e2e}--\ref{tab:system_ablation} and Fig.~\ref{fig:rate_sweep_10db} report separate evaluation runs, including separate AWGN references in Tables~\ref{tab:e2e} and~\ref{tab:rayleigh}.

\subsection{Carrier Compactness (Non-Rate-Matched)}
Table~\ref{tab:payload} summarizes the payload study on 2,000 Cityscapes images. Layouts use lossless portable network graphics (PNG); RAG features and edges use 16-bit floating point (fp16)+\texttt{zlib}. Compression is lossless after serialization, but fp16 can introduce quantization. The RAG payload is smaller on every image, but pre-channel bits are not channel uses; Section~\ref{sec:ratematched} provides the channel-level evidence.

\begin{table}[!t]
\caption{Pre-channel \emph{carrier compactness} on 2,000 Cityscapes images. PNG and zlib are lossless; fp16 can quantize. Not rate-matched.}
\label{tab:payload}
\centering

\begin{tabular}{lcc}
\toprule
 & Layout (PNG) & RAG (fp16+zlib) \\
\midrule
Mean bits & 45\,969.8 & 8\,170.5 \\
Median bits & 45\,256.0 & 8\,200.0 \\
RAG smaller on & \multicolumn{2}{c}{2,000 / 2,000 images} \\
\bottomrule
\end{tabular}
\end{table}

\subsection{Transmission-Cost Accounting}
\label{sec:cost}
Table~\ref{tab:tx_cost} itemizes the transmitted object. Its 2,000-image carrier study and Table~\ref{tab:fixedadaptive}'s 500-image validation study use different populations, so their mean node counts and channel uses need not match. Because transmission is analog, the primary rate metric is channel uses ($M\cdot d$ real-valued symbols); the listed 16-bit size is serialization only, not transmitted.

\begin{table}[!t]
\caption{Transmission cost on 2,000 Cityscapes images; channel uses are real symbols, 16-bit sizes are storage only, and $\overline{M}$ is the mean retained-node count.}
\label{tab:tx_cost}
\centering
 
\begin{tabular}{ccccc}
\toprule
\shortstack{SNR\\(dB)} & $\overline{M}$ &
\shortstack{Ch.\\uses} &
\shortstack{Bandwidth\\ratio $\rho$} &
\shortstack{Storage\\(16-bit, kbit)} \\
\midrule
0  & 27.2 & 1\,742 & $3.3 \times 10^{-3}$ & 27.9 \\
5  & 34.0 & 2\,175 & $4.1 \times 10^{-3}$ & 34.8 \\
10 & 40.8 & 2\,610 & $5.0 \times 10^{-3}$ & 41.8 \\
15 & 47.6 & 3\,046 & $5.8 \times 10^{-3}$ & 48.7 \\
20 & 54.4 & 3\,480 & $6.6 \times 10^{-3}$ & 55.7 \\
\bottomrule
\end{tabular}
\end{table}

Channel uses count only $Md$ real-valued embedding scalars; node count, ordering, coordinates or regions, and connectivity are excluded metadata. Original pixels and full segmentation maps are not transmitted. The embedding-only ratio $\rho\approx3$--$7\times10^{-3}$ is below DeepJSCC's $0.021$, but metadata exclusions and resolution differences preclude a total-cost claim. Practical digital transmission adds quantization, topology signaling, framing, and possibly channel coding.

\subsection{End-to-End Comparison with Baselines}
Table~\ref{tab:e2e} reports the end-to-end comparison on 500 validation images with four channel-noise realizations per image. Because carrier rates differ, the 10~dB sweep in Section~\ref{sec:ratematched} is the primary carrier-level evidence. Across 0--15~dB, CA-RAG reports higher mIoU than the tested baselines in this protocol; relative to the same-decoder layout baseline, it also gives lower LPIPS at 0, 10, and 15~dB and lower FID at every tested SNR. DeepJSCC is not resolution- or decoder-matched, so it is treated as a JSCC-style reference.

\begin{table}[!t]
\caption{End-to-end AWGN comparison; 500 images; four channel-noise realizations per image. Layout is the GESCO-based baseline. \textbf{Bold}: best per SNR; parentheses: standard deviations.}
\label{tab:e2e}
\centering
 
\setlength{\tabcolsep}{2pt}
\begin{tabular}{@{}llcccc@{}}
\toprule
 & SNR (dB) & 0 & 5 & 10 & 15 \\
\midrule
\multirow{3}{*}{LPIPS$\downarrow$}
  & DeepJSCC       & 0.673 (.021) & 0.663 (.022) & 0.657 (.024) & 0.653 (.025) \\
  & Layout & 0.587 (.024) & \textbf{0.561 (.025)} & 0.561 (.025) & 0.559 (.025) \\
  &  Ours  & \textbf{0.571 (.025)} & 0.563 (.024) & \textbf{0.555 (.025)} & \textbf{0.551 (.023)} \\
\midrule
\multirow{3}{*}{mIoU$\uparrow$}
  & DeepJSCC       & 0.014 (.011) & 0.016 (.011) & 0.017 (.011) & 0.017 (.011) \\
  & Layout & 0.142 (.037) & 0.257 (.054) & 0.255 (.055) & 0.251 (.049) \\
  &  Ours  & \textbf{0.233 (.062)} & \textbf{0.264 (.069)} & \textbf{0.294 (.071)} & \textbf{0.312 (.070)} \\
\midrule
\multirow{3}{*}{FID$\downarrow$}
  & DeepJSCC       & 425.2 & 389.6 & 360.6 & 350.9 \\
  & Layout & 110.3 & 81.8 & 81.4 & 81.8 \\
  &  Ours  & \textbf{74.8} & \textbf{74.1} & \textbf{74.8} & \textbf{74.1} \\
\bottomrule
\end{tabular}
\end{table}

The GESCO-inspired layout baseline and CA-RAG share the ControlNet+diffusion decoder, but the conditioning signals differ; this partially controls for the decoder rather than fully isolating the carrier.

\begin{figure*}[!tb]
\centering
\includegraphics[width=0.8\textwidth]
{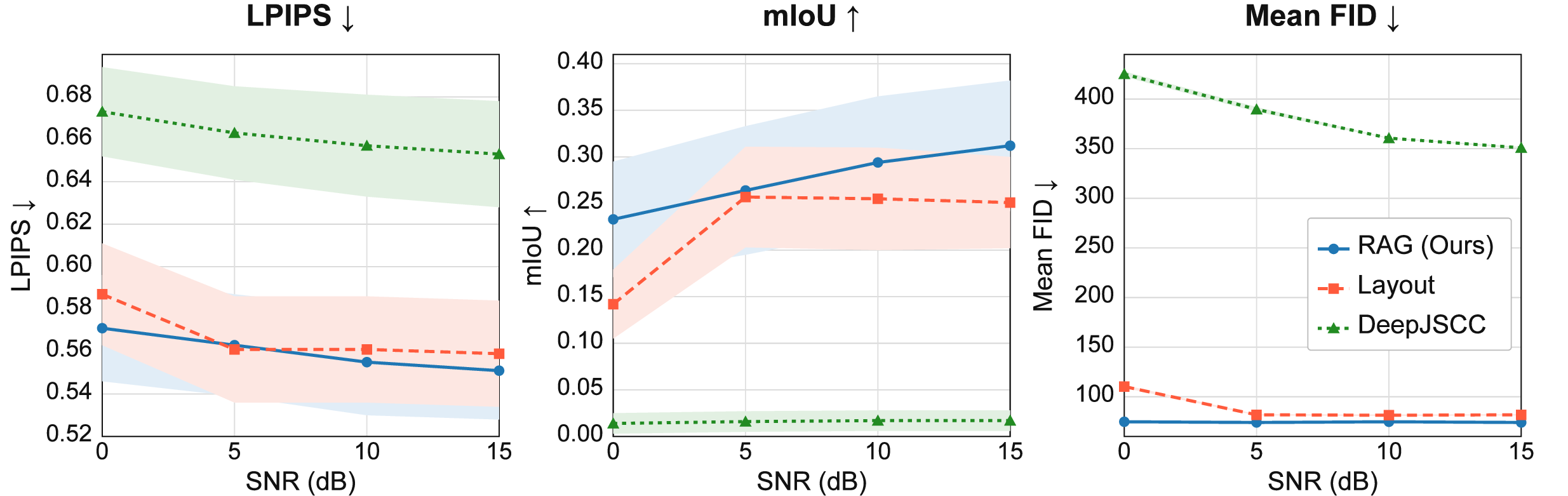}
\caption{End-to-end AWGN comparison over SNR.}
\label{fig:main_results}
\end{figure*}

\subsection{Rate--Quality Comparison at 10~dB}
\label{sec:ratematched}
\begin{figure*}[!tb]
\centering
\includegraphics[width=0.55\textwidth]
{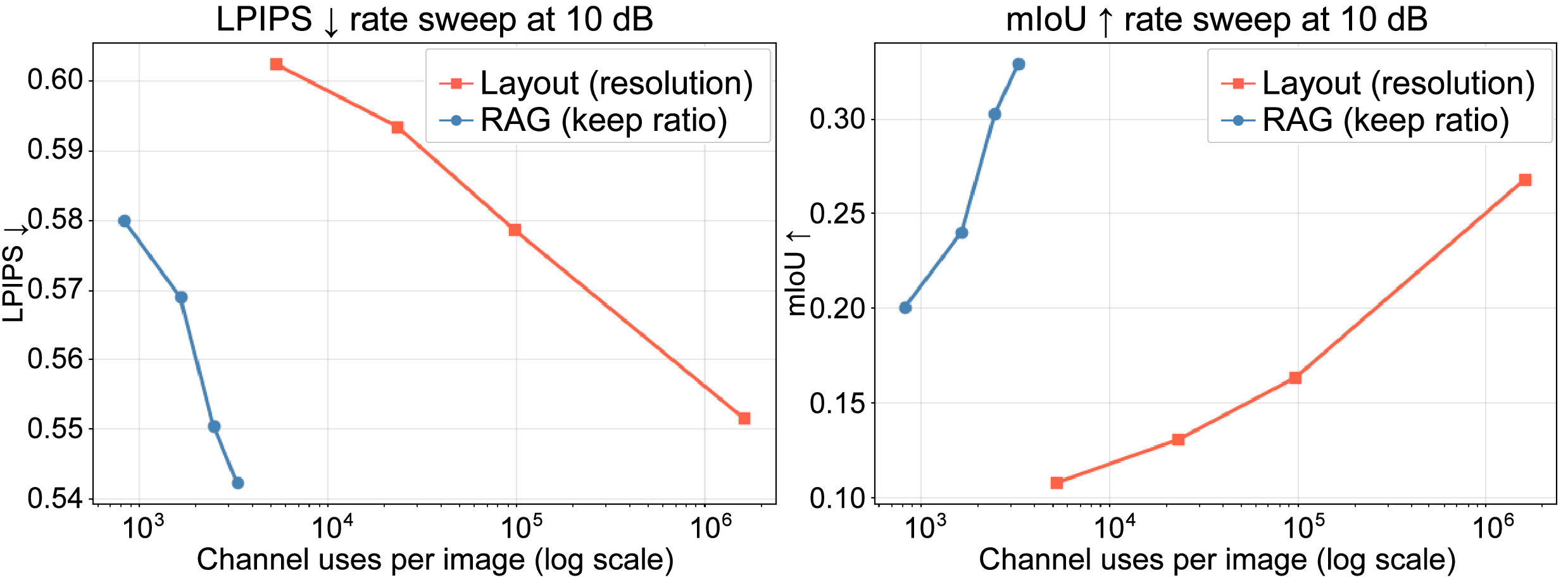}
\caption{Carrier-level rate--quality sweep at 10~dB, excluding CA-RAG metadata. RAG has lower LPIPS and higher mIoU than the layout baseline.}
\label{fig:rate_sweep_10db}
\end{figure*}

Because the methods operate at different rates (Table~\ref{tab:tx_cost} and \cite{8723589,Grassucci_2026}), the evaluation includes a 10~dB rate sweep. RAG keep-ratios are $r\in\{0.25, 0.50, 0.75, 1.00\}$; layout resolutions are $\{16{\times}32, 32{\times}64, 64{\times}128, 256{\times}512\}$. Figure~\ref{fig:rate_sweep_10db} plots LPIPS/mIoU versus channel uses. These sweep budgets are separate from the default adaptive-CGS rates in Table~\ref{tab:tx_cost}.

In this sweep, the evaluated layout points are less favorable than the RAG points in both metrics, even at substantially higher channel-use budgets. At RAG's full-budget point ($\approx\!3.3\times 10^3$ channel uses), the proposed method reaches LPIPS = 0.542 and mIoU = 0.329; the lowest-rate layout baseline ($5.3\times 10^3$ channel uses) reaches LPIPS = 0.602, mIoU = 0.108, and the highest-rate layout baseline ($\approx\!1.6\times 10^6$ channel uses) reaches LPIPS = 0.552, mIoU = 0.268. The two rate ranges do not overlap exactly, so this is an empirical observation rather than a formal Pareto-dominance claim.

\begin{figure*}[!tb]
\centering
\includegraphics[width=0.7\textwidth]
{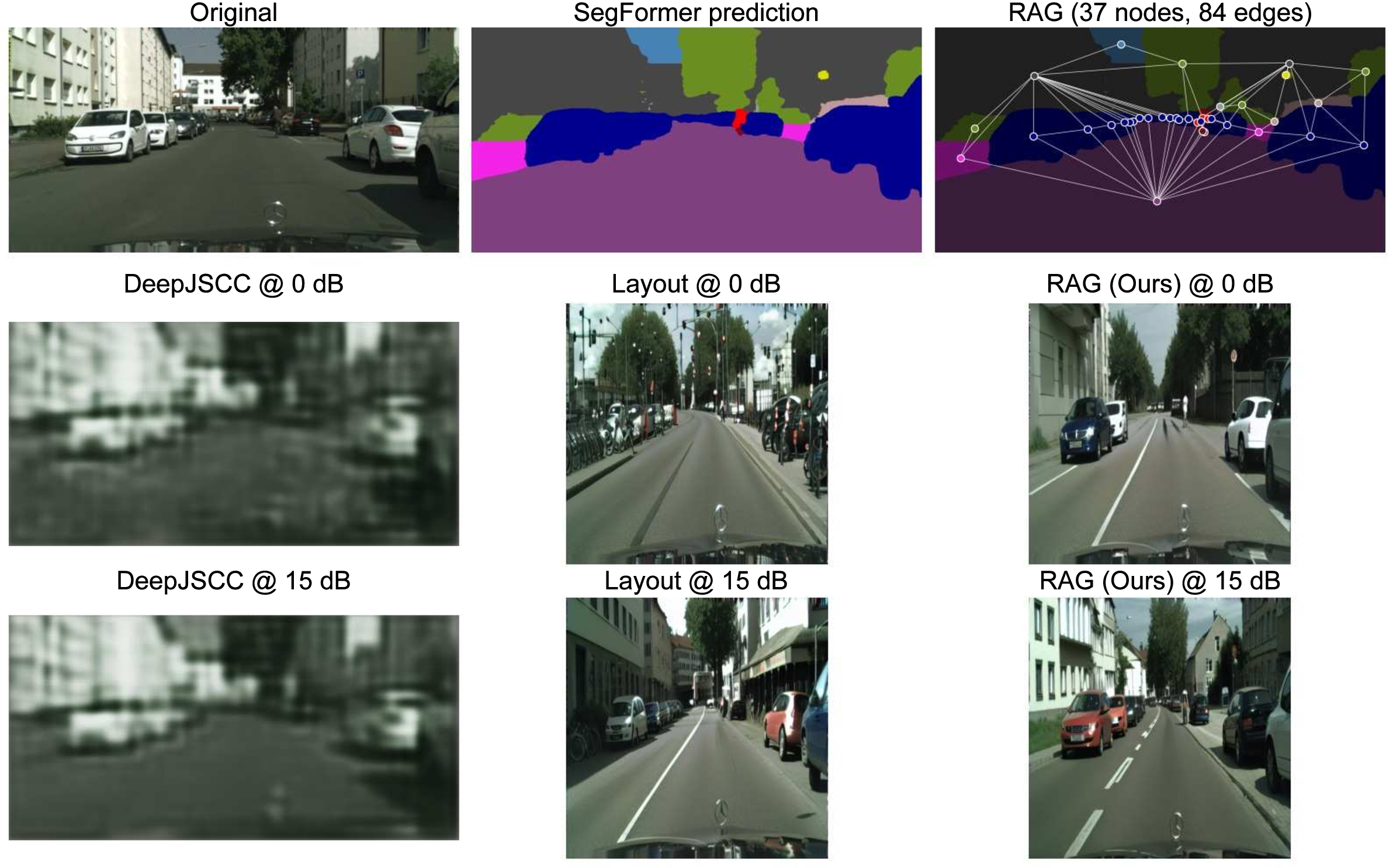}
\caption{Qualitative comparison at 0/15~dB. Bandwidth ratios: DeepJSCC $\rho\!\approx\!0.021$, layout carrier $\rho\!\approx\!3.0$, and CA-RAG $\rho\!=\!3.3\times10^{-3}$/$5.8\!\times\!10^{-3}$.}
\label{fig:qualitative}
\end{figure*}

\subsection{Fixed Budget and Rate Sweep}
Table~\ref{tab:fixedadaptive} compares adaptive CGS with a fixed-max variant. Fixed-max yields lower LPIPS and higher mIoU at low SNR by transmitting up to roughly $2\times$ more symbols, confirming that CGS is a \emph{rate controller} rather than a quality booster.

\begin{table}[!t]
\caption{Adaptive CGS versus fixed-max graph budget on 500 validation images with four channel-noise realizations per image.}
\label{tab:fixedadaptive}
\centering
 
\setlength{\tabcolsep}{3pt}

\begin{tabular}{c cccc cccc}
\toprule
& \multicolumn{4}{c}{Adaptive CGS (ours)} & \multicolumn{4}{c}{Fixed-max (no adaptation)} \\
\cmidrule(lr){2-5}\cmidrule(lr){6-9}
SNR & LPIPS & mIoU & Rate & Nodes & LPIPS & mIoU & Rate & Nodes \\
\midrule
0  & 0.571 & 0.233 & 1\,774 & 27.7 & 0.550 & 0.300 & 3\,548 & 55.4 \\
5  & 0.563 & 0.264 & 2\,219 & 34.7 & 0.547 & 0.312 & 3\,548 & 55.4 \\
10 & 0.555 & 0.294 & 2\,661 & 41.6 & 0.548 & 0.321 & 3\,548 & 55.4 \\
15 & 0.551 & 0.312 & 3\,105 & 48.5 & 0.548 & 0.318 & 3\,548 & 55.4 \\
\bottomrule
\end{tabular}

\end{table}

Figure~\ref{fig:rate_sweep_10db} shows the same rate trend: increasing the retained node budget is associated with lower LPIPS and higher mIoU.

\subsection{Rayleigh Fading (Idealized)}
\label{sec:rayleigh}
Table~\ref{tab:rayleigh} evaluates idealized Rayleigh fading with perfect-CSI zero-forcing on 500 validation images with four channel realizations per image. We test \emph{node-block} fading (per-embedding draws) and \emph{graph-block} fading (one draw per image) against an AWGN baseline.

\begin{table}[!t]
\caption{Idealized Rayleigh fading vs.\ AWGN on 500 validation images with four channel realizations per image.}
\label{tab:rayleigh}
\centering
 
\setlength{\tabcolsep}{3.5pt}
\begin{tabular}{llcccc}
\toprule
Channel & Mode & \multicolumn{2}{c}{SNR = 0 dB} & \multicolumn{2}{c}{SNR = 15 dB} \\
\cmidrule(lr){3-4}\cmidrule(lr){5-6}
 & & LPIPS$\downarrow$ & mIoU$\uparrow$ & LPIPS$\downarrow$ & mIoU$\uparrow$ \\
\midrule
AWGN     & --           & 0.571 & 0.237 & 0.550 & 0.309 \\
Rayleigh & node-block   & 0.572 & 0.229 & 0.550 & 0.310 \\
Rayleigh & graph-block  & 0.570 & 0.233 & 0.549 & 0.307 \\
\bottomrule
\end{tabular}
\end{table}

Under idealized perfect-CSI equalization, both fading modes stay within 0.003 LPIPS and 0.009 mIoU of AWGN; deep fades can amplify noise.

\subsection{Channel Robustness: Structured Node Erasure}
\label{sec:dropout}
\begin{figure}[!tb]
\centering
\includegraphics[width=0.48\textwidth]
{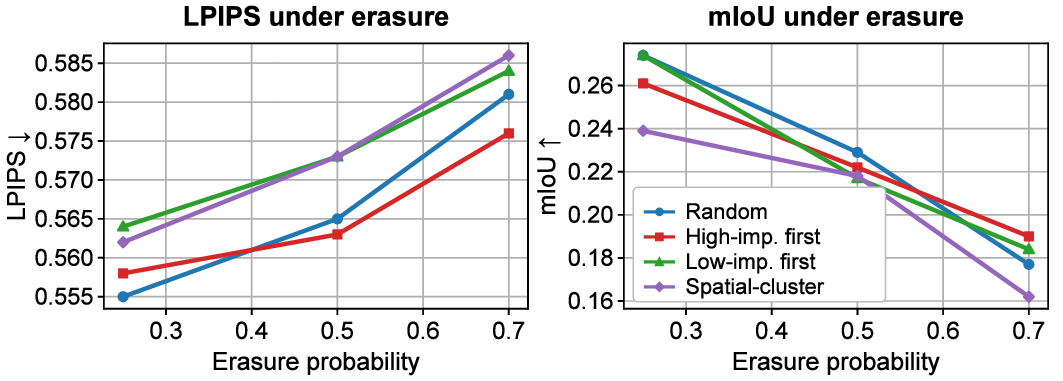}
\caption{Structured node erasure at AWGN 10~dB on 500 validation images with four channel-noise realizations per image.}
\label{fig:dropout}
\end{figure}

Node erasure after AWGN is a synthetic robustness stress test. At SNR = 10~dB on 500 validation images with four channel-noise realizations per image, we test random, high-importance-first, low-importance-first, and spatial-cluster erasure; surviving nodes are fed to SBP, while dropped nodes are removed from $\tilde{\mathcal{G}}$. Figure~\ref{fig:dropout} merges the LPIPS and mIoU trends over erasure probability $p$.

All four strategies show gradual degradation: at 70\% erasure, LPIPS rises by at most 0.037 and mIoU drops by at most 0.144 relative to no erasure ($p=0$: LPIPS $\approx 0.549$, mIoU $\approx 0.299$). Spatial-cluster erasure causes the largest high-$p$ mIoU drop, but SBP's contribution is not isolated.

\begin{figure}[!tb]
\centering
\includegraphics[width=0.48\textwidth]
{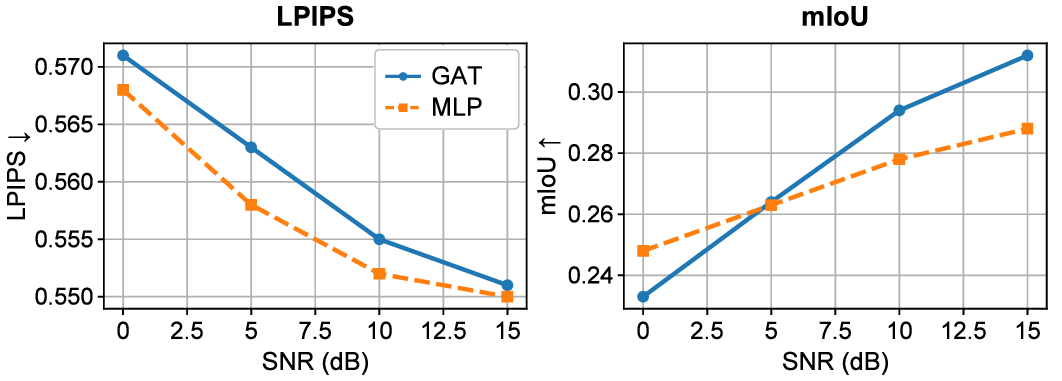}
\caption{Encoder ablation on 500 validation images with four channel-noise realizations per image: GAT vs.\ per-node MLP.}
\label{fig:encoder_ablation}
\end{figure}

\subsection{CGS Parameter Sensitivity}
\label{sec:cgs_sensitivity}
Table~\ref{tab:cgs_sens} sweeps $(r_{\min}, r_{\max})$ around the default $(0.5,1.0)$. LPIPS varies by at most 0.016 and mIoU by at most 0.059, tracking the node count rather than parameter instability.

\begin{table}[!t]
\caption{CGS $(r_{\min}, r_{\max})$ sensitivity at 10~dB on 500 validation images with four channel-noise realizations per image. The $^\ast$ denotes the default configuration.}
\label{tab:cgs_sens}
\centering
 
\begin{tabular}{lccc}
\toprule
$(r_{\min}, r_{\max})$ & $\overline{M}$ & LPIPS$\downarrow$ & mIoU$\uparrow$ \\
\midrule
$(0.30, 0.80)$          & 28.5 & 0.561 & 0.259 \\
$(0.30, 1.00)$          & 33.6 & 0.556 & 0.284 \\
$(0.40, 0.90)$          & 33.6 & 0.556 & 0.278 \\
$(0.50, 1.00)^\ast$     & 38.8 & 0.550 & 0.301 \\
$(0.60, 1.00)$          & 41.5 & 0.545 & 0.318 \\
$(0.70, 1.00)$          & 44.0 & 0.545 & 0.313 \\
\midrule
Range (max $-$ min)     &      & 0.016 & 0.059 \\
\bottomrule
\end{tabular}
\end{table}

\subsection{Ablations: CGS, SBP, Encoder, and Learned Budget}
\label{sec:ablations}
Table~\ref{tab:system_ablation} isolates CGS and SBP at 10~dB. Removing CGS sends the full graph at higher rate and gives lower LPIPS and higher mIoU. Removing SBP changes LPIPS/mIoU by 0.001/0.001 with CGS and 0.004/0.014 without CGS. Its AWGN effect is modest; erasure tests do not isolate SBP.

\begin{table}[!t]
\caption{CGS/SBP ablation (10~dB; 500 images; four channel-noise realizations per image. ``No~CGS'' transmits the full graph.}
\label{tab:system_ablation}
\centering
 
\begin{tabular}{lcc}
\toprule
Variant & LPIPS$\downarrow$ & mIoU$\uparrow$ \\
\midrule
Full system         & 0.551 (.024) & 0.290 (.063) \\
No CGS (full graph) & \textbf{0.546 (.026)} & \textbf{0.319 (.074)} \\
No SBP              & 0.550 (.025) & 0.291 (.069) \\
No CGS + No SBP     & 0.550 (.027) & 0.305 (.062) \\
\bottomrule
\end{tabular}
\end{table}

Figure~\ref{fig:encoder_ablation} compares the GAT encoder with a per-node MLP. The MLP slightly lowers LPIPS, but GAT gives higher mIoU at 5--15~dB and a higher mean mIoU (0.276 vs.\ 0.269); message passing may help preserve semantic structure.

A lightweight learned-budget predictor collapsed in the larger evaluation (LPIPS $\approx 0.78$, mIoU $\approx 0.07$ across 0--15~dB), for reasons not isolated experimentally. A jointly trained continuous-relaxation selector is a natural next step.

\subsection{Qualitative Reconstruction}

Figure~\ref{fig:qualitative} shows reconstructions at 0 and 15~dB. DeepJSCC recovers little identifiable scene content; the layout baseline is coherent but less detailed; and the RAG reconstruction shows sharper object boundaries. Because the GESCO-inspired layout baseline and CA-RAG use generative diffusion receivers, we evaluate semantic consistency and perceptual similarity rather than pixel-level peak signal-to-noise ratio (PSNR).


\subsection{Discussion, Limitations, and Complexity}

Limitations include segmentation errors, uncounted packetization/topology overhead, diffusion hallucinations, and the unmatched DeepJSCC reference. Shared decoding partly controls comparisons. Fading assumes perfect CSI and floored zero-forcing; imperfect CSI, outage, and frequency selectivity remain open. CGS trades quality for fewer channel uses. With fixed architectures and SBP iterations, encoder and SBP inference cost $O(N+|\mathcal{E}|)$ ($|\mathcal{E}|\lesssim110$), excluding graph preprocessing and CGS selection/merging. Diffusion dominates inference. Findings remain specific to Cityscapes.

\section{Conclusion}

CA-RAG provides segmentation-derived, rate-controllable graph carriers for JSCC-style image communication. It explicitly represents regions and relations, adapting the transmission budget to scene complexity and channel conditions. Under the reported Cityscapes protocol, it reports higher mIoU than the tested baselines, comparable LPIPS to a same-decoder layout carrier, and favorable 10~dB rate--quality behavior over the evaluated operating points. Idealized Rayleigh and structured node-erasure stress tests assess protocol-specific robustness. 


\bibliographystyle{IEEEtran}
\bibliography{refs}

@article{qin2021semanticcommunicationsprincipleschallenges,
  title={Semantic communications: {P}rinciples and challenges},
  author={Qin, Zhijin and Tao, Xiaoming and Lu, Jianhua and Tong, Wen and Li, Geoffrey Ye},
  journal={arXiv preprint arXiv:2201.01389},
  year={2021}
}

@article{ismail2026semantic,
  title={Semantic Communications in the {THz} Band},
  author={Ismail, Fatima and Sarieddeen, Hadi and Fahs, Jihad},
  journal={arXiv preprint arXiv:2607.07455},
  year={2026}
}

@article{alhaj2026signdeepsc,
  title={{SignDeepSC: A} Semantic Signature-based Approach for Robust Semantic Communication},
  author={Alhaj, Khalil and Tajeddine, Razane and Sarieddeen, Hadi},
  journal={arXiv preprint arXiv:2607.25676},
  year={2026}
}

@ARTICLE{9398576,
  author={Xie, Huiqiang and Qin, Zhijin and Li, Geoffrey Ye and Juang, Biing-Hwang},
  journal={IEEE Transactions on Signal Processing}, 
  title={Deep Learning Enabled Semantic Communication Systems}, 
  year={2021},
  volume={69},
  number={},
  pages={2663-2675},
  doi={10.1109/TSP.2021.3071210}}

@ARTICLE{8723589,
  author={Bourtsoulatze, Eirina and Burth Kurka, David and Gündüz, Deniz},
  journal={IEEE Transactions on Cognitive Communications and Networking}, 
  title={Deep Joint Source-Channel Coding for Wireless Image Transmission}, 
  year={2019},
  volume={5},
  number={3},
  pages={567-579},
  doi={10.1109/TCCN.2019.2919300}}

@article{Grassucci_2026,
   title={Generative Semantic Communication: Diffusion Models Beyond Bit Recovery},
   volume={12},
   ISSN={2372-2045},
   DOI={10.1109/tccn.2026.3689849},
   journal={IEEE Transactions on Cognitive Communications and Networking},
   publisher={Institute of Electrical and Electronics Engineers (IEEE)},
   author={Grassucci, Eleonora and Barbarossa, Sergio and Comminiello, Danilo},
   year={2026},
   pages={8171--8185} }

@ARTICLE{841950,
  author={Tremeau, A. and Colantoni, P.},
  journal={IEEE Transactions on Image Processing}, 
  title={Regions adjacency graph applied to color image segmentation}, 
  year={2000},
  volume={9},
  number={4},
  pages={735-744},
  doi={10.1109/83.841950}}

@article{velickovic2018graphattentionnetworks,
  title={Graph attention networks},
  author={Veli{\v{c}}kovi{\'c}, Petar and Cucurull, Guillem and Casanova, Arantxa and Romero, Adriana and Lio, Pietro and Bengio, Yoshua},
  journal={arXiv preprint arXiv:1710.10903},
  year={2017}
}

@incollection{779343.779352,
author = {Yedidia, Jonathan S. and Freeman, William T. and Weiss, Yair},
title = {Understanding belief propagation and its generalizations},
year = {2003},
isbn = {1558608117},
booktitle = {Exploring Artificial Intelligence in the New Millennium},
pages = {239--269},
numpages = {31}
}

@inproceedings{NEURIPS2021_64f1f27b,
  author    = {Enze Xie and Wenhai Wang and Zhiding Yu and Anima Anandkumar
               and Jose M. Alvarez and Ping Luo},
  title     = {{SegFormer: S}imple and Efficient Design for Semantic Segmentation
               with Transformers},
  booktitle = {Advances in Neural Information Processing Systems},
  volume    = {34},
  pages     = {12077--12090},
  year      = {2021}
}

@inproceedings{Cordts_2016_CVPR,
  title={The cityscapes dataset for semantic urban scene understanding},
  author={Cordts, Marius and Omran, Mohamed and Ramos, Sebastian and Rehfeld, Timo and Enzweiler, Markus and Benenson, Rodrigo and Franke, Uwe and Roth, Stefan and Schiele, Bernt},
  booktitle={Proceedings of the IEEE conference on computer vision and pattern recognition (CVPR)},
  pages={3213--3223},
  year={2016}
}

@INPROCEEDINGS{10377881,
  author={Zhang, Lvmin and Rao, Anyi and Agrawala, Maneesh},
  booktitle={2023 IEEE/CVF International Conference on Computer Vision (ICCV)}, 
  title={Adding Conditional Control to Text-to-Image Diffusion Models}, 
  year={2023},
  volume={},
  number={},
  pages={3813-3824},
  doi={10.1109/ICCV51070.2023.00355}}

@INPROCEEDINGS{9878449,
  author={Rombach, Robin and Blattmann, Andreas and Lorenz, Dominik and Esser, Patrick and Ommer, Björn},
  booktitle={2022 IEEE/CVF Conference on Computer Vision and Pattern Recognition (CVPR)}, 
  title={High-Resolution Image Synthesis with Latent Diffusion Models}, 
  year={2022},
  volume={},
  number={},
  pages={10674-10685},
  doi={10.1109/CVPR52688.2022.01042}}

\end{document}